\documentclass[twocolumn,showpacs,preprintnumbers,amsmath,amssymb]{revtex4}

\usepackage{graphicx}

\usepackage{dcolumn}

\usepackage{bm}

\newcommand{\ket}[1]{|#1\rangle}

\begin{document}

\title{Relaxation and dephasing in a flux qubit}

\author{P. Bertet$^1$, I. Chiorescu$^{1*}$, G. Burkard$^{2}$, K. Semba$^{1,3}$, C. J. P. M. Harmans$^1$,D.P. DiVincenzo$^{2}$, J. E. Mooij$^1$}

\affiliation{$^1$Quantum Transport Group, Kavli Institute of Nanoscience, Delft University of Technology, Lorentzweg $1$,$2628CJ$, Delft, The Netherlands \\
$^2$ IBM T.J. Watson Research Center, P.O. Box 218, Yorktown
Heights, NY 10598, USA \\
$^3$ NTT Basic Research Laboratories, Atsugi-shi, Kanagawa
243-0198, Japan}

\begin{abstract}

We report detailed measurements of the relaxation and dephasing
time in a flux-qubit measured by a switching DC SQUID. We studied
their dependence on the two important circuit bias parameters~:
the externally applied magnetic flux and the bias current through
the SQUID in two samples. We demonstrate two complementary
strategies to protect the qubit from these decoherence sources.
One consists in biasing the qubit so that its resonance frequency
is stationary with respect to the control parameters ({\it optimal
point}) ; the second consists in {\it decoupling} the qubit from
current noise by chosing a proper bias current through the SQUID.
At the decoupled optimal point, we measured long spin-echo decay
times of up to $4 \mu s$.

\end{abstract}


\maketitle

A long-standing problem for the use of superconducting circuits as
quantum bits (qubits) in a quantum computer
\cite{Makhlin02,qubits,Vion02,Chiorescu03} is their relatively
short dephasing time compared to the requirements of many-qubit
quantum computation. Dephasing is due to the coupling of the
qubit's degrees of freedom with the many fluctuating uncontrolled
ones commonly denoted as the environment
\cite{Makhlin02,decoherence}. From the perspective of quantum
information, it is crucial to quantitatively identify the various
dephasing sources and to find strategies to overcome these, either
by reducing the amount of fluctuations or by protecting the qubit
against it. An important step in this direction has been
accomplished in \cite{Vion02}. The authors showed that dephasing
can be significantly reduced by biasing the qubit at an {\it
optimal} point where its resonance frequency is stationary with
respect to its control parameters - in that case, gate voltage and
magnetic flux.

In this letter we report detailed measurements of the relaxation
and dephasing times in a flux-qubit as a function of its bias
parameters for two different samples. Our measurements allow us to
identify certain dephasing mechanisms and quantify their effect on
the qubit. We find that energy relaxation is dominated by
spontaneous emission towards the measuring circuit impedance.
Dephasing is mainly caused by noise in the external magnetic flux
biasing the qubit, thermal fluctuations of our measuring circuit,
and low-frequency noise originating from microscopic degrees of
freedom, probably causing critical current noise in the qubit
junctions. We moreover demonstrate strategies to efficiently fight
each of these noise sources.

Our flux-qubit consists of a micron-size superconducting loop
intersected with three Josephson junctions \cite{Mooij99}. When
the total phase across the three junctions $\gamma_Q$ is close to
$\pi$, the loop has two low-energy eigenstates (ground state
$\ket{0}$ and excited state $\ket{1}$) well separated from the
higher-energy ones, which can thus be used as a qubit
\cite{Caspar00,Chiorescu03}. The flux-qubit is characterized by
two parameters : the minimum energy separation $\Delta$ between
$\ket{0}$ and $\ket{1}$, and the persistent current $I_p$. Around
$\gamma_Q=\pi$, the energy separation between these two levels
depends on $\gamma_Q$ and can be written as $E_1-E_0 \equiv h f_Q
= h \sqrt{\Delta ^2 + \epsilon ^2 }$, where $\epsilon \equiv
(I_p/e) (\gamma_Q-\pi)/(2\pi)$. The qubit is inductively coupled
to a SQUID detector (with a coupling constant $M$), which is
biased at a current $I_b$. The phase drop $\gamma_Q$ has two
origins : the magnetic flux threading the qubit loop $\Phi_x$, and
the currents in the SQUID loop which depend on $I_b$. Thus, we can
write $\epsilon=\eta(\Phi_x)+\lambda(I_b)$.

The coupling of $\epsilon$ to fluctuating sources leads to
decoherence.  Noise in the magnetic flux $\Phi_x$ or in the bias
current $I_b$ induces fluctuations of the qubit frequency $f_Q$
and thus dephasing. A first strategy to protect the qubit from
decoherence consists in biasing it at $\epsilon=0$ so that $df_Q/d
\epsilon=0$. This is the {\it optimal point} strategy, which was
first invented and demonstrated in \cite{Vion02}. An additional
possibility is to {\it decouple} the external noise from the
variable $\epsilon$, by canceling the sensitivity coefficients $d
\eta / d \Phi_x$ and $d \lambda / d I_b$. The flux noise can not
be decoupled since $d \eta/d \Phi_x= 2 I_p/h$ is constant. As we
will show below, the bias current noise can be decoupled by
biasing the SQUID at a current $I_b^*$ such that $d \lambda/d I_b
(I_b^*)=0$, which is the {\it decoupling} condition. At the
decoupled optimal point, ($\epsilon=0$ and $I_b=I_b^*$) we expect
that the qubit quantum coherence is best preserved, since the
qubit is sensitive to flux noise to second order, and to bias
current noise to fourth order. We also note that a strong
dependence of the dephasing time on the bias current would be
clear experimental evidence that current noise, and not flux
noise, is the factor limiting the quantum coherence.

\begin{figure}
\resizebox{.48\textwidth}{!}{\includegraphics{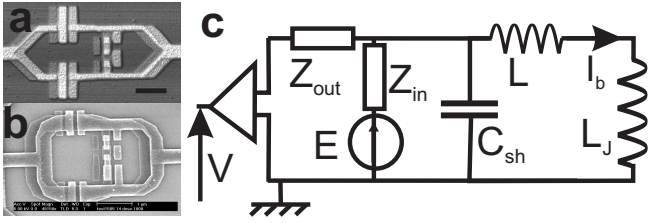}}
\caption{(a) Atomic Force Micrograph of sample A. The flux qubit
is the small loop containing three Josephson junctions in a row ;
the SQUID is constituted by the outer loop containing the two
large junctions. The bar indicates a $1 \mu m$ length. (b)
Scanning Electron Micrograph of sample B. Note that the qubit loop
contains a fourth junction, $3$ times larger than the other ones.
(c) Electrical model of the measuring circuit. The SQUID,
represented by its Josephson inductance $L_J$, is shunted by an
on-chip capacitor $C_{sh}$ through superconducting lines of
inductance $L$ (all on-chip). It is current-biased by a waveform
generator delivering a voltage $E$ across an impedance $Z_{in}$ ;
the voltage across the SQUID is connected to the input of a
room-temperature preamplifier through an impedance $Z_{out}$.
$Z_{in}$ and $Z_{out}$ include on-chip gold resistors.
 \label{fig1}}
\end{figure}

In the two samples, shown in figures \ref{fig1}a and \ref{fig1}b,
the qubit loop is merged with its measuring SQUID. The dependence
of $\epsilon$ on the bias current $I_b$ arises from the way this
bias current redistributes in the SQUID and eventually generates a
phase shift across the qubit junctions via the superconducting
line shared by the qubit and the SQUID. The detailed configuration
of the shared line is related to the specific fabrication process.
We use 2-angle shadow evaporation so that the lines consist
effectively of 2 layers. This induces a large asymmetry in the
coupling \cite{Guido04} if the qubit loop contains an odd number
of junctions. In this article, we compare the results obtained for
a three- (sample A) and a four- (sample B) junction qubit (see
figure \ref{fig1}a and b), and demonstrate that such asymmetry can
be removed by using an even number of junctions in the qubit loop
\cite{Guido04}. A model for the qubit electromagnetic environment
in both samples is shown in figure 1c. The SQUID is modeled by its
Josephson inductance $L_J$ shunted by a capacitor $C_{sh}$ via
superconducting lines of inductance $L$. It is connected to the
output voltage of our waveform generator $E$ via an impedance
$Z_{in}$, and to the input of a room-temperature amplifier through
an impedance $Z_{out}$. It thus forms a harmonic oscillator, the
plasma mode, of frequency $\omega_p=(\sqrt{(L+L_J)C_{sh}})^{-1}$
and quality factor $Q=\omega_p C_{sh} Re(Z)(\omega_p)$, to which
the qubit is strongly coupled \cite{Chiorescu04} (we note
$Z=Z_{in}//Z_{out}$). Here is a list of the parameters for our two
samples : for sample A, $I_p=270nA$, $\Delta=5.85GHz$, $M=20pH$,
$L_J=80pH$, $L=170pH$, $C=12pF$, $Z(0)=1.4k \Omega$ ; for sample
B, $I_p=240nA$, $\Delta=5.5GHz$, $M=6.5pH$, $L_J=380pH$, $L=80pH$,
$L_J=350pH$, $C=5.5pF$, $Z(0)=9k\Omega$. We note that the main
difference between sample A and B, apart from the number of
junctions in the qubit loop, is the value of $M$ and of the
low-frequency impedance $Z(0)$.

\begin{figure}
\resizebox{.48\textwidth}{!}{\includegraphics{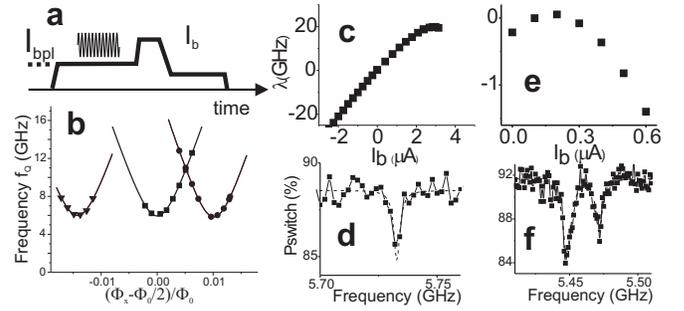}}
\caption{(a) Principle of the spectroscopy experiments : a bias
current pulse of amplitude $I_b{bpl}$ (lower than the SQUID
critical current) is applied to the sample while a microwave (MW)
pulse probes the qubit resonance frequency. The qubit state is
finally measured by a short bias current pulse as discussed in
\cite{Chiorescu03}. (b) Typical spectroscopy curves for three
values of $I_{bpl}$ measured with sample A (from left to right,
$I_{bpl}=-2.25,0,2 \mu A$). The solid curves are fits to the data.
(c,e) Curves $\lambda(I_b)$ deduced from the spectroscopy curves
as explained from the text for sample A (c) and sample B (e). The
decoupling condition is satisified at $I_b^*=2.9 \pm 0.1 \mu A$
for sample A (black arrow in the figure) and $I_b^*=180 \pm 20 nA$
for sample B. (d,f) Qubit line at the decoupled optimal point for
sample A (d) and B (f).}
 \label{fig2}
\end{figure}

We measured our sample parameters and $\lambda (I_b)$ by studying
the dependence of the qubit Larmor frequency on both the external
flux $\Phi_x$ and the bias current $I_b$. We performed
spectroscopy by applying a $500ns$ microwave pulse of variable
frequency, and measuring the SQUID switching probability with a
short subsequent DC current pulse \cite{Chiorescu03} for different
values of $\Phi_x$. We added a $1 \mu s$ plateau at the value
$I_{bpl}$ in order to adjust the bias current through the SQUID
{\it during} the application of the microwave pulse. The complete
pulse sequence is depicted in figure \ref{fig2}a. In figure
\ref{fig2}b the measured qubit resonance frequency for sample A is
shown as a function of the external flux $\Phi_x$ for three
different values of $I_{bpl}$. We observe that for each value of
the bias current, a specific value of external flux
$\Phi_x^{(0)}(I_{bpl})$ realizes the optimal point condition.

We fitted all the curves with the formula
$f_Q=\sqrt{\Delta^2+[\lambda (I_b)+2I_p (\Phi_x-\Phi_0/2)/h]^2}$
for different values of $I_b$. The obtained curves $\lambda (I_b)$
are shown in figure \ref{fig2}c and \ref{fig2}e for both samples.
The decoupling occurs at $I_b^*=2.9 \pm 0.1 \mu A$ for sample A
and at $I_b^*=180 \pm 20 nA$ for sample B. Note that although the
SQUID critical current is similar in both samples, the decoupling
current is much closer to $0$ in sample B due to the presence of
the fourth junction wich restores the symmetry of the coupling
\cite{Guido04}. We biased our qubit at the decoupled optimal point
by setting $I_b=I_b^*$ and $\Phi_x=\Phi_x^{(0)}(I_b^*)$. The qubit
line shape under these conditions is shown in figure \ref{fig2}d
for sample A and \ref{fig2}f for sample B. For sample A, we could
fit it with a Lorentzian of width $w=3.1 \pm 0.5MHz$ (FWHM). This
width yields a dephasing time $T_2=1/\pi w \simeq 100ns$
consistent with the Ramsey fringe measurements as discussed below.
For sample B, the line was split, due to the action of a strongly
coupled two-level fluctuator. We fitted it by the sum of two
Lorentzians of widths $7$ and $6MHz$. We note that in addition to
the fluctuator responsible for the splitting of the line, the
value of the qubit frequency at the optimal point $\Delta$
exhibited occasional jumps of around $100MHz$. Also the width of
the line changed significantly in time. This indicates that
dephasing was probably dominated by some low-frequency noise due
to one or more strongly coupled microscopic fluctuators, likely
generating critical current noise. We stress that we had no
evidence for such instabilities with sample A.

\begin{figure}
\resizebox{.48\textwidth}{!}{\includegraphics{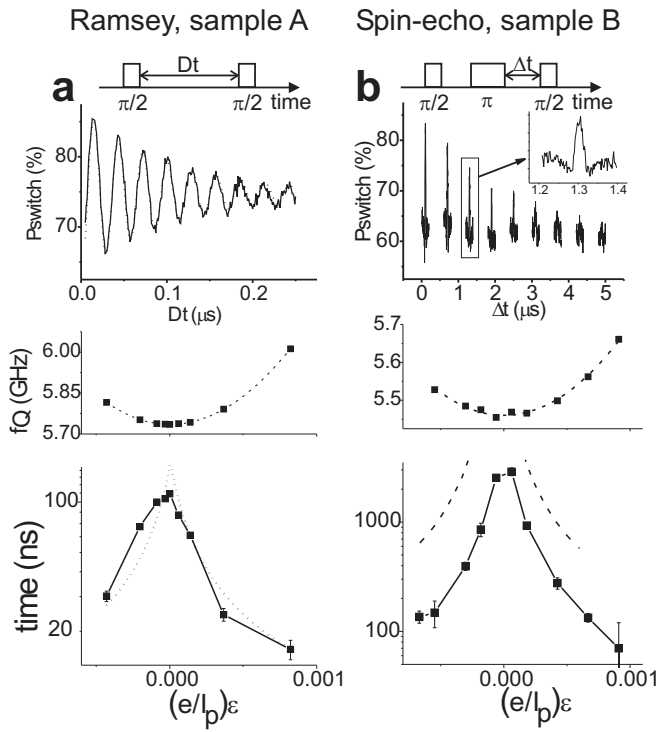}}
\caption{(a) Ramsey fringe signal measured with sample A. From top
to bottom : - temporal sequence of microwave pulses corresponding
to a Ramsey experiment. - Ramsey signal at the optimal decoupled
point. - Qubit resonance frequency as a function of $\epsilon$. -
Dephasing time $T_2$ as a function of $\epsilon$ around the
optimal point (full squares, the lines are a guide to the eye),
and fit to the data (dotted curve) assuming that dephasing was
caused by $1/f$ flux noise $S_{\Phi_x}=3 \cdot 10^{-12}/f [
\Phi_0^2/Hz]$. (b) Spin-echo signal measured with sample B. From
top to bottom : Temporal sequence of microwave pulses
corresponding to a spin-echo experiment.
 - Spin-echo signal at the optimal decoupled point. - Qubit resonant
frequency as a function of $\epsilon$. - (full squares) :
Spin-echo time $T_{echo}$ as a function of $\epsilon$ (the lines
are a guide to the eye). (dashed curve) : calculated dephasing
from thermal fluctuations of the photon number in the SQUID plasma
mode.
 \label{fig3}}
\end{figure}

We first studied the dependence of the dephasing time as a
function of $\epsilon$ while keeping $I_b=I_b^*$. For sample A, we
measured Ramsey fringes \cite{Vion02,Chiorescu03} by applying a
sequence of microwave pulses as schematized in figure \ref{fig3}a
for each value of $\Phi_x$. The Ramsey fringes measured at the
decoupled optimal point are shown. They decay exponentially with a
time constant $T_2$. Figure \ref{fig3}a (bottom) shows the
dependence of $T_2$ with $\Phi_x$. The dephasing time exhibits a
sharp maximum $T_2=120ns$ at the optimal point $\epsilon=0$ for
which $d f_Q / d \epsilon=0$ as expected.

To account for the rapid degradation of the dephasing time when
$\epsilon \neq 0$, we first evaluated the effect of the thermal
fluctuations in the measuring circuit on the qubit coherence time.
Thermal fluctuations of the photon number in the plasma mode cause
fluctuations of the qubit resonance frequency and thus dephasing.
It has been shown \cite{Thorwart04} that treating the thermal
fluctuations of the plasma mode as a weak classical perturbation
leads to a strong underestimate because of the neglect of quantum
correlations between the qubit and the oscillator. Instead, we
numerically integrated the master equation for the joint density
matrix of the ``qubit-plasma mode" system \cite{Haroche}. The
qubit density matrix is obtained at the end of the calculation by
tracing over the plasma mode degrees of freedom. The evolution of
its off-diagonal element yields the dephasing time. For these
calculations, we assumed a quality factor $Q=100$ for sample $A$
and $Q=150$ for sample B, and an effective temperature $T=70mK$ in
agreement with additional measurements \cite{Chiorescu04} ; all
the other parameters of the model are directly obtained from
experimental data. For sample A, we found that thermal
fluctuations have a negligible effect when $I_b=I_b^*$ ; we thus
believe that flux-noise is responsible for rapid degradation of
the dephasing time when $\epsilon \neq 0$. Assuming that the
flux-noise power spectrum has a frequency dependence given by
$S_{\Phi_x}=A/|f|$ which is consistent with noise measurements
found in the literature \cite{Clarke}, we can use the data from
figure \ref{fig3}a to evaluate $A$. With calculations similar to
\cite{Nakamuraecho}, we find that $A=3 \pm 1.5 \cdot 10^{-12}
\Phi_0^2$ gives a good agreement (dotted line in figure
\ref{fig3}a). Such a level of noise is comparable to the lowest
values reported in SQUID measurements \cite{Clarke,Foglietti}.

As could be expected from the lineshape shown in figure
\ref{fig2}f, the Ramsey fringe signal measured with sample B had a
non-exponential damping so that it was impossible to measure
$T_2$. To circumvent the low-frequency noise mentioned above, we
used a spin-echo type sequence of microwave pulses shown in figure
\ref{fig3}b (top), as demonstrated in the case of low-frequency
charge noise \cite{Nakamuraecho,Collin04}. The results are shown
in figure \ref{fig3}b at the decoupled optimal point, by a set of
curves corresponding to different delays between the two $\pi/2$
pulses. We fitted each curve by a gaussian multiplied by a sine
curve. Then we fitted the decay of the echo amplitude as a
function of the delay between the two $\pi/2$ pulses with an
exponential of time constant $T_{echo}$. At the decoupled optimal
point, we measured $T_{echo}=3.9 \pm 0.1 \mu s$. We stress that
these results represent a significant improvement over previously
reported coherence times in superconducting qubits. We studied the
dependence of $T_{echo}$ as a function of $\epsilon$ (figure
\ref{fig3}b bottom, full squares) for $I_b=I_b^*$. Again we found
a sharp maximum at $\epsilon=0$. For sample B, thermal
fluctuations in the plasma mode account qualitatively for the
experimental data (dashed curve in figure \ref{fig3}b bottom).

\begin{figure}
\resizebox{.48\textwidth}{!}{\includegraphics{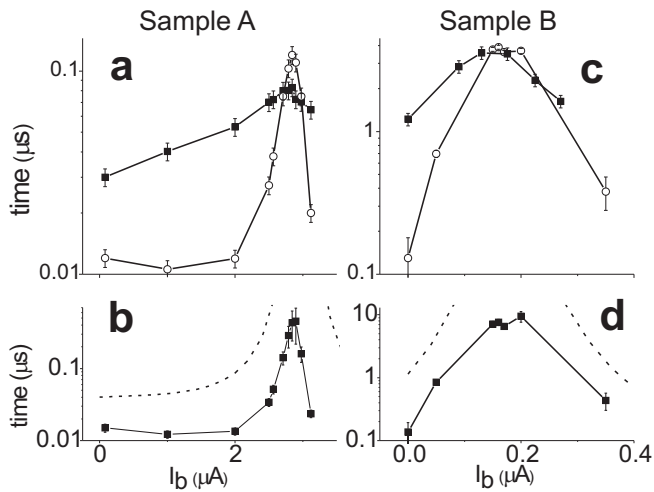}}
\caption{(a) Relaxation ($T_1$, black squares) and dephasing
($T_2$, empty circles) times at the optimal point ($\epsilon=0$)
as a function of the bias current $I_b$ for sample A. (b) Pure
dephasing time $T_\phi$ (full squares) as a function of $I_b$. The
dashed line is the result of a simulation taking into account the
thermal fluctuations of the plasma mode. (c) Relaxation ($T_1$,
black squares) and echo ($T_2$, empty circles) times at the
optimal point ($\epsilon=0$) as a function of the bias current
$I_b$ for sample B. (d) Pure dephasing component of the echo time
$T_\phi^{echo}$ (full squares) and calculated effect of the
thermal fluctuations in the plasma mode (dashed line).
 \label{fig4}}
\end{figure}

We finally studied the bias current $I_b$ dependence of the
dephasing and echo times $T_2$ and $T_{echo}$ together with the
energy relaxation time $T_1$ at $\epsilon=0$. The results are
shown in figure \ref{fig4}a and c for samples A and B
respectively. All these curves exhibit a clear maximum at
$I_b=I_b^*$. This indicates that at $I_b \neq I_b^*$ both
relaxation and dephasing are limited by coupling to the measuring
circuit. In particular, the fact that in sample B $T_1$ is
strongly reduced away from $I_b^*$ is a clear indication that
energy relaxation occurs by spontaneous emission towards the
circuit impedance seen by the qubit. A weaker dependence is
observed for sample A, which could indicate that additional
environmental modes at the qubit frequency are involved. The
dephasing time $T_2$ measured in sample A is strongly dependent on
$I_b$. This indicates that dephasing at the optimal point is
limited by noise in the bias current. For both samples, the
dephasing time measured at the optimal decoupled point is similar
to or larger (sample A) than the relaxation time, so that
dephasing was partly limited by relaxation. To quantify the pure
dephasing contribution, we calculated $T_\phi$ defined as
$T_\phi^{-1} \equiv (T_2)^{-1}-(2T_1)^{-1}$ for sample A (full
square curve in figure \ref{fig4}b) and calculated similarly
$T_\phi^{echo}$ for sample B (full square curve in figure
\ref{fig4}b). Our calculations taking into account the thermal
fluctuations in the plasma mode are shown as the dashed curve in
figures \ref{fig4}b and d. They are in qualitative agreement with
the data, although systematically overestimating the dephasing
time by a factor typically $5$ compared to the measurements.

In conclusion, we presented detailed measurements of the
relaxation and dephasing times as a function of bias parameters
for two flux-qubit samples. We showed that the {\it optimal point}
concept already demonstrated for the quantronium circuit
\cite{Vion02} is also valid for the flux-qubit design. Making use
of the SQUID geometry of our detector, we could moreover {\it
decouple} the qubit from current fluctuations by biasing the SQUID
at a specific current $I_b^*$. We showed that adding a fourth
junction to the qubit loop enhances the symmetry of the coupling,
thus lowering the value of $I_b^*$. We showed that low-frequency
noise limits the dephasing time, but that spin-echo techniques
provide a powerful tool to fight it. We observed remarkably long
decay times of the echo signal of $4 \mu s$, limited by
relaxation. We provided quantitative evidence that at the optimal
point dephasing is induced by the thermal fluctuations of the
photon number in the plasma mode of our SQUID detector. These
results indicate that long coherence times can be achieved with
flux qubits.

We thank Y. Nakamura, D. Est\`{e}ve, D. Vion, M. Grifoni for
fruitful discussions. This work was supported by the Dutch
Foundation for Fundamental Research on Matter (FOM), the E.U.
Marie Curie and SQUBIT grants, and the U.S. Army Research Office.

\end{document}